\documentclass[aps,prd,amssymb,showpacs,floatfix,twocolumn,superscriptaddress,nofootinbib]{revtex4-2}
\usepackage{amsmath,amsfonts,graphics,epsfig}
\usepackage[T1]{fontenc}
\usepackage[utf8]{inputenc}
\usepackage[english]{babel}

\begin{document}

\title{Double integral representation in the invariant masses of a composite system and
nonperturbative calculation of the asymptotics of light-meson electromagnetic form factors}

	\author{A.F.~Krutov}
        \email{a$_$krutov@rambler.ru} \affiliation{Samara State Technical
		University, 443100 Samara, Russia}
	\author{E.V.~Razueva}
	\email{evgenia.razueva@yandex.ru} \affiliation{Lyceum 135 -
		basic school of the Russian Academy of Sciences, 443077 Samara, Russia}
	\author{V.E.~Troitsky}
	\email{troitsky@theory.sinp.msu.ru} \affiliation{D.V.~Skobeltsyn Institute
		of Nuclear Physics,\\
		M.V.~Lomonosov Moscow State University, Moscow 119991, Russia}
	\date{\today}

\begin{abstract}
We obtain nonperturbative asymptotic expansions of the electromagnetic form factors of the
$\pi$ and $\rho$ mesons at large spacelike momentum transfer, $Q^2\to\infty$.
The calculation is based on a double integral representation in the invariant masses of the
composite system. The model parameters are kept at the values fixed in the authors' earlier
description of meson electroweak properties. We show that, in the limit of vanishing constituent-quark mass, with the constituent quarks
taken to be pointlike, the asymptotic ratio of the longitudinally polarized $\rho$-meson to
pion form factors in the helicity basis is consistent with the prediction of perturbative QCD
(pQCD) within the adopted parameter ranges.
\end{abstract}

	\maketitle

\section{Introduction}
Electromagnetic form factors of composite quark systems, and of light mesons in particular,
at large momentum transfer have been the subject of intensive theoretical and experimental
study for several decades (see, e.g., Ref.~\cite{GrK23}). Electromagnetic form factors
(EMFFs), together with gravitational form factors (GFFs), provide information on hadron
structure and on the spatial distributions of charge, magnetic moment, and other physical
quantities \cite{BuE23, ChC24, XuD24, ChC25, QuF25}. At large momentum transfer, they
probe hadronic structure at short distances.

At asymptotically large momentum transfer, predictions are available from perturbative QCD
(pQCD) \cite{FaJ79, LeB79, EfR80, LeB80} and from constituent-counting rules \cite{MaM72}. 
Determining the onset of the perturbative regime is therefore an
important problem. Interest in this momentum-transfer region is also driven by the programs 
of ongoing and planned accelerator experiments \cite{And21, Geo22, Chr22, Bur23}.
Experimental measurements of meson form factors have already reached
large values of $Q^2$; for example, the charged-kaon form factor in the timelike region has
been measured up to about $50\,(\mathrm{GeV}/c)^2$ \cite{Lee15}. Nevertheless, present pQCD
calculations do not yet provide a complete description of the available data \cite{QuF25}.

For example, Refs.~\cite{ChC24, ChC25} calculated subleading terms in the pion and kaon
charge form factors within collinear factorization at next-to-next-to-leading order (NNLO).
Agreement with experiment was found to depend appreciably on nonperturbative input,
including the parameters of the leading-twist distribution amplitude and its Gegenbauer
moments. This illustrates that, at momentum transfers currently accessible experimentally,
the perturbative description remains strongly dependent on nonperturbative information and,
by itself, does not yet yield a satisfactory description of all available data.

A natural requirement on a nonperturbative hadron model is that it reproduce the known QCD
asymptotics, allow a transition from the soft to the hard regime, and at the same time
describe data at low and intermediate $Q^2$. For spacelike form factors, a useful framework
is a double integral representation (DIR) in the invariant masses of the composite system
\cite{AnS87, TrS69, KiT75}. Different choices of the building blocks entering the DIR
correspond to different models of the constituent interaction. The application of such a DIR
to the electromagnetic form factors of the $\pi$ and $\rho$ mesons at asymptotically large
momentum transfer is the subject of the present work.

We use the authors' nonperturbative relativistic approach to composite systems formulated in
the instant form (IF) of relativistic dynamics \cite{KrT02, KrT03, KrT09}. Dirac introduced
three forms of relativistic dynamics \cite{Dir49}; their formulations and characteristic 
features are discussed in detail in Refs.~\cite{LeS78, KeP91, Coe92, KrT09}.
Besides the IF, the point form (PF) \cite{Gom14, HaA19} and light-front dynamics (LF) 
\cite{CaG96, Pol23, RaX26} are widely
used. 
The IF formulation employed here has previously been applied to electroweak and gravitational
properties of composite hadrons \cite{KrT02, KrT03, KrT01, KrT09prc, TrT13, KrP15, KrP16,
KrP18, KrT22, KrT23, KrT25}.

In particular, this approach predicted the pion charge form factor over a broad range of
$Q^2$ \cite{KrT01}; subsequent measurements were found to be consistent with that prediction
\cite{KrT09prc}. In calculations of the pion GFFs \cite{KrT22, KrT23, KrT25}, most model
parameters were inherited from the earlier description of the pion electromagnetic
properties \cite{KrT01}, while the constituent-quark $D$ term was fixed separately from
gravitational information. In the present calculation all relevant parameters have already
been fixed, so the asymptotic behavior is obtained without any additional fit of the model
parameters.

The same framework has also been applied to the electromagnetic properties of the $\rho$
meson at low momentum transfer \cite{KrP15, KrP16, KrP18}. 
In particular,
Ref.~\cite{KrP16} used the measured leptonic decay constant to obtain
$\langle r_\rho^2\rangle=(0.56\pm0.04)\,\mathrm{fm}^2$. 
The electromagnetic structure of the $\rho$ meson,
and more generally of spin-1 systems, has also been studied in other relativistic schemes,
including LF and PF approaches \cite{Bra18, Mel19, HaA21}; phenomenological information on the
$\rho$-meson magnetic moment has also been extracted from BABAR data \cite{RoT24}.

For the pion, it was previously shown that in the zero constituent-mass limit the asymptotic
charge form factor agrees with the pQCD prediction \cite{TrT13}. Here we derive asymptotic
expansions of the pion and $\rho$-meson EMFFs, including subleading terms, using a general
method for asymptotic expansion of double integrals of a special type \cite{KrT08}. Since
pQCD relates the asymptotic form factors of the pion and the longitudinally polarized
$\rho$ meson \cite{LeB80}, reproducing an analogous relation within the present approach
provides a test of whether the same framework can connect the soft and hard regimes.

A further motivation comes from the continuing discussion of which form of relativistic
dynamics is most useful for nonperturbative QCD problems \cite{LuS25}. In this context,
deriving asymptotic form-factor relations in the IF and confronting them with pQCD
predictions is of independent interest and may provide an additional argument for the
applicability of the IF to nonperturbative problems.

The paper is organized as follows. Section~\ref{sec: Sec 2} briefly describes the DIR used for
the electromagnetic form factors. Their asymptotic expansions are derived in
Sec.~\ref{sec: Sec 3}. In Sec.~\ref{sec: Sec 4} the results are compared with pQCD
predictions. The main conclusions are summarized in Sec.~\ref{sec: Sec 5}.

\section{Double integral representation for the electromagnetic form factors of the $\pi$ and $\rho$ mesons in instant-form dynamics}
\label{sec: Sec 2}

We analyze the pion and $\rho$-meson EMFFs within the relativistic nonperturbative framework
developed in Refs.~\cite{KrT02, KrT03}. In this approach, form factors of a two-particle
composite system are written as dispersion-type double integrals over the invariant masses of
the initial and final states. We first briefly discuss the degree of generality of this
representation.

Dispersion representations are particularly convenient for studying form-factor asymptotics.
For example, Ref.~\cite{QuF25} used such a representation to analyze the pion form factor in
the timelike region and to investigate how restrictions on individual components of the
dispersion integral affect its asymptotic behavior.

For the spacelike region considered here, it is natural to employ a DIR whose general form was
derived, for example, in Ref.~\cite{AnS87}. There the representation follows from the
triangle diagram with two-particle intermediate states and its characteristic logarithmic
(triangle-diagram) singularity. Assuming that the EMFF of the composite system inherits this
singularity, one obtains, for equal constituent masses,
\begin{equation}
	F(Q^2) = \int_{4M^2}^\infty\frac{ds}{2\pi}\frac{ds'}{2\pi}\,\frac{G(s)}{s - M_c^2}
	D_M(s,Q^2,s')\frac{G(s')}{s' - M_c^2}\;,
\label{FAnis}
\end{equation}
where $q$ is the four-momentum transfer, $Q^2=-q^2>0$, $M$ is the constituent mass,
$M_c$ is the mass of the composite system, $G(s)$ is the vertex function, and $D_M$
describes the transition induced by a virtual photon from an on-shell two-particle state of
invariant mass $\sqrt{s'}$ to an on-shell state of invariant mass $\sqrt{s}$.

The function $D_M$ is constructed under the assumption that the virtual photon couples to only
one constituent; consequently, Eq.~(\ref{FAnis}) corresponds to a relativistic impulse
approximation. The vertex function $G(s)$ may contain the so-called anomalous singularities,
whose structure depends on the chosen model of the constituent interaction. In the
terminology of Ref.~\cite{AnS87}, the vertex function is a prototype of a wave function.
Indeed, in the nonrelativistic limit the factor $G(s)(s-M_c^2)^{-1}$ reduces to the wave
function of the relative constituent motion.

In the earlier studies \cite{TrS69, KiT75}, which preceded Ref.~\cite{AnS87}, a representation
analogous to Eq.~(\ref{FAnis}) was derived for composite systems whose constituents possess
scattering states. Using the relation between the discontinuities of the electromagnetic
current matrix element across the kinematic and anomalous cuts and the physical, on-shell
$S$ matrix \cite{TrS69}, one obtains for a two-particle composite-system form factor a
representation analogous to Eq.~(\ref{FAnis}). For a simple composite system with vanishing
total spin it has the structure \cite{KiT75}
$$
	F(Q^2) = \frac{\Gamma^2}{2\pi\,i\,B'(M_c^2)}
	\int_{4M^2}^\infty\,dsds'\frac{\Delta(s)}{s - M_c^2}
$$	
\begin{equation}	
	\times g_0(s,Q^2,s')\frac{\Delta(s')}{s' - M_c^2}\;,
	\label{FTrS}
\end{equation}
where $\Gamma^2$ is a normalization constant fixed by the static limit of the form factor at
$Q^2\to0$, $\Delta(s)=B(s+i0)-B(s-i0)$, $B(s)$ is the relativistic Jost function,
$B'(M_c^2)$ is its derivative at the bound-state pole, and $g_0(s,Q^2,s')$ is the free
two-particle form factor. The latter describes the electromagnetic properties of two
mutually noninteracting constituents carrying the quantum numbers of the original composite
system, i.e., in a state with specified orbital angular momentum and total spin.

The function $g_0(s,Q^2,s')$ in Eq.~(\ref{FTrS}) is a regular generalized function
(distribution) on a space of test functions (see, e.g., Refs.~\cite{BoL90, LoC17, KrT02,
KrT03}). In particular, its static limit $Q^2\to0$ exists only in the weak sense. The free
two-particle form factor can be evaluated, although the calculation is rather lengthy, using
relativistic kinematics \cite{KrT02, KrP18} together with the general relativistic
parametrization of matrix elements of local operators \cite{ChS63, CoL20}. The expressions
for two-particle systems with the pion and $\rho$-meson quantum numbers are collected in the
Appendix. Equation~(\ref{FTrS}) also shows explicitly that the composite-system form factor
can be expressed through the constituent scattering phases encoded in the Jost functions.

Thus, the DIR in Eq.~(\ref{FAnis}) is sufficiently general for the EMFFs of two-particle
composite systems, while different choices of the vertex function implement different models
of the constituent interaction.

An important observation made in Refs.~\cite{KrT02, KrT03} is that the vertex functions in
Eqs.~(\ref{FAnis}) and (\ref{FTrS}) can be related to wave functions beyond the
nonrelativistic limit. They may be represented in terms of relativistic wave functions in the
sense of Dirac forms of relativistic dynamics. We now summarize the particular construction
used below for meson EMFFs.

The DIR developed in Refs.~\cite{KrT02, KrT03} was formulated within instant-form
relativistic quantum mechanics (RQM). The approach has previously been applied to the
electromagnetic and gravitational structure of composite particles, including the $\rho$
meson \cite{KrT02, KrT03, KrP15, KrP16, KrP18}. Pion GFFs are described by analogous DIRs
with the same approximations; most model parameters are taken from the electromagnetic
pion analysis, while the constituent-quark $D$ term is additionally fixed from gravitational
information \cite{KrT22, KrT23, KrT25}.

A defining feature of Dirac's relativistic quantum dynamics is that the interparticle
interaction is incorporated into the algebra of Poincar\'e-group generators through
interaction-dependent generators. In the instant-form construction used here, the
interaction in the composite system is introduced by adding an interaction operator to the
mass operator of the free system, as in the conventional IF formulation of RQM. The
constituents are assumed to remain on their mass shells, and the wave function of the
interacting system is defined as an eigenfunction of a complete commuting set of operators.
For IF RQM we use
	\begin{equation}
		{\hat M}_I^2\;
		(\hbox{or}\;\hat M_I = \hat M_0 + \hat V)\;,\quad
		{\hat J}^2\;,\quad \hat J_3\;,\quad \hat {\vec P}\;,
		\label{complete}
	\end{equation}
where $\hat M_0$ denotes the free-system mass operator, with eigenvalue $\sqrt{s}$;
$\hat V$ is the interaction operator; $\hat M_I$ is the interacting mass operator;
$\hat J^2$ and $\hat J_3$ are the total-angular-momentum squared and its $z$ projection;
and $\hat{\vec P}$ is the total-momentum operator. The components of $\hat{\vec J}$ are
constructed in the standard way as invariants formed from the components of the
Pauli--Lubanski four-pseudovector operator (see, e.g., Ref.~\cite{Nov76}).

In IF RQM the operators $\hat J^2$, $\hat J_3$, and $\hat{\vec P}$ coincide with their
counterparts for the noninteracting composite system; only $\hat M_I^2$ (or $\hat M_I$)
contains the interaction. The first three operators can therefore be diagonalized in a
two-particle basis in which the center-of-mass motion is separated and the orbital angular
momentum and total spin are fixed. It follows that the internal wave function is an
eigenfunction of $\hat M_I^2$ (or $\hat M_I$). For the $\pi$ and $\rho$ mesons with
equal-mass constituents it is written as (see, e.g., Refs.~\cite{KrT23, KrT02})
	$$
	\varphi^{(\pi,\rho)}(s(k)) = \sqrt[4]{s}\,k\,u^{(\pi,\rho)}(k)\;,\quad s = 4(k^2 + M^2)\;,
	$$
	\begin{equation}
		\int\,\left[u^{(\pi,\rho)}(k)\right]^2\,k^2\,dk = 1\;,
		\label{phi}
	\end{equation}
where $u^{(\pi,\rho)}(k)$ specifies the model momentum-space wave function of the
quark-antiquark pair and $M_u=M_{\bar d}=M$ is the common constituent mass. Exact isospin
symmetry is assumed.

Substituting the wave function (\ref{phi}) for the vertex function in the DIR
(\ref{FAnis}) gives the following representations for the pion and $\rho$-meson EMFFs:
	\begin{equation}
	G^{(\pi)}_{10}(Q^2) =	\int\,d\sqrt{s}\,d\sqrt{s'}\,
	\varphi^{(\pi)}(s)\,G^{(00)}_{10}(s,Q^2,s')\,
	\varphi^{(\pi)}(s')\;,
	\label{int ds=Gpi0}
\end{equation}
$$
G^{(\rho)}_{ij}(Q^2) =
\int\,d\sqrt{s}\,d\sqrt{s'}\,
\varphi^{(\rho)}(s)\,G^{(01)}_{ij}(s,Q^2,s')\,
\varphi^{(\rho)}(s')\;,
$$
\begin{equation}
	(ij) = (10)\;,\;(12)\;,\;(30)\;,
	\label{int ds=Grho0}
\end{equation}
where $G^{(0J)}_{ij}$, $J=0,1$, are the free two-particle electromagnetic form factors.
The indices $(10)$, $(12)$, and $(30)$ correspond to the charge, quadrupole, and magnetic
form factors, respectively. Explicit expressions for $G^{(00)}_{10}$ and
$G^{(01)}_{ij}$ are given in the Appendix.

The DIRs (\ref{int ds=Gpi0}) and (\ref{int ds=Grho0}) rely on the same physical assumptions
as Eq.~(\ref{FAnis}). All these representations are based on the impulse approximation.
In deriving Eqs.~(\ref{int ds=Gpi0}) and (\ref{int ds=Grho0}), however, we use a modified
impulse approximation (MIA) which, unlike the conventional impulse approximation, preserves
Lorentz covariance and electromagnetic-current conservation \cite{KrT02}.
The free two-particle form factors entering these equations have the
same physical meaning as $g_0$ in Eq.~(\ref{FTrS}): they depend on the invariant masses of
the initial and final two-particle states, while $Q^2$ appears as a parameter. Mathematically,
they are regular generalized functions (distributions), represented by functionals defined by
the two-dimensional invariant-mass integrals in Eqs.~(\ref{int ds=Gpi0}) and
(\ref{int ds=Grho0}) \cite{KrT02}. In particular, their static limits as $Q^2\to0$ must be
understood in the weak sense.

We next relate the form factors appearing in Eqs.~(\ref{int ds=Gpi0}) and
(\ref{int ds=Grho0}) to the conventional form factors. Since only the invariant parts of the
matrix elements are needed, it is sufficient to establish the relation in a convenient frame.
For the $\rho$ meson, the electromagnetic-current matrix element in the Breit frame can be
written as \cite{ArC80, BrH92}
$$
\langle\vec p_\rho\,,m_\rho|j_\mu(0)|\vec p_\rho\,'\,,m'_\rho\rangle =
G^\mu(Q^2)\;,
$$
$$
G^0(Q^2) =
2p_{\rho0}\left\{(\vec\xi\,'\vec\xi\,^*)\,G^{(\rho)}_C(Q^2)\right. +
$$
$$
\left. \left[(\vec\xi\,^*\vec Q)(\vec\xi\,'\vec Q) -
\frac{1}{3}Q^2
(\vec\xi\,'\vec\xi\,^*)\right] \,\frac{G^{(\rho)}_Q(Q^2)}{2M_\rho^2}\right\}\;,
$$
\begin{equation}
	\vec G(Q^2) = \frac{p_{\rho
			0}}{M_\rho}\left[\vec\xi'\,(\vec\xi\,^*\vec Q) -
	\vec\xi\,^*(\vec\xi\,'\vec Q)\right]\,G^{(\rho)}_M(Q^2)\;,
	\label{Grho}
\end{equation}
where $G^{(\rho)}_C$, $G^{(\rho)}_Q$, and $G^{(\rho)}_M$ are the usual Sachs charge,
quadrupole, and magnetic form factors of the $\rho$ meson, respectively, and
$$
q^\mu = (p_\rho - p'_\rho)^\mu = (0\;,\;\vec Q)\;,
$$
$$
p_\rho^\mu = (p_{\rho0}\;,\;\frac{1}{2}\vec Q)\;,\quad
p'_\rho\,^\mu = (p_{\rho0}\;,\;-\frac{1}{2}\vec Q)\;,
$$
$$
p_{\rho0} = \sqrt{M_\rho^2 + \frac{1}{4}Q^2}\;,\quad \vec Q =
(0\;,\;0\;,\;Q)\;.
$$
$$
\xi^\mu(\pm 1) = \frac{1}{\sqrt{2}}(0\;,\;\mp
1\;,\;-\,i\;,\;0)\;,
$$
$$
\xi^\mu(0) = (0\;,\;0\;,\;0\;,\;1)\;.
$$
The label of the polarization four-vector $\xi^\mu$ specifies the projection of the total
angular momentum.

The DIR form factors are related to the Sachs form factors in Eq.~(\ref{Grho}) by
\cite{KrT03}
$$
G^{(\pi)}_C(Q^2) = G^{(\pi)}_{10}(Q^2)\;,\quad
G^{(\rho)}_C(Q^2) = G^{(\rho)}_{10}(Q^2)\;,
$$
$$
	\quad G^{(\rho)}_Q(Q^2) =
	\frac{2\,M_\rho^2}{Q^2}\,G^{(\rho)}_{12}(Q^2)\;,
$$
\begin{equation}
	G^{(\rho)}_M(Q^2) = -\,M_\rho\,G^{(\rho)}_{30}(Q^2)\;.
	\label{G=f}
\end{equation}

Equations~(\ref{int ds=Gpi0}) and (\ref{int ds=Grho0}) are double integrals of a special
type. Their asymptotic expansions for $Q^2\to\infty$ are derived in the next section.

\section{Asymptotic behavior of pion and $\rho$-meson electromagnetic form factors}
	\label{sec: Sec 3}

In this section we derive the large-momentum-transfer asymptotic expansions of the pion and
$\rho$-meson EMFFs defined by Eqs.~(\ref{int ds=Gpi0}) and
(\ref{int ds=Grho0}) within instant-form relativistic dynamics.

As seen from Eqs.~(\ref{int ds=Gpi0}) and (\ref{int ds=Grho0}), the pion and $\rho$-meson
form factors in our approach are represented by double integrals of a special type. The
boundary of the integration domain is fixed by the step function $\vartheta(s,Q^2,s')$
entering the free two-particle form factors (see the Appendix), and therefore depends on the
asymptotic-expansion parameter $Q^2$. Theorems determining the form of asymptotic expansions
for integrals of this class, together with the corresponding expansion formulas, were derived
in Ref.~\cite{KrT08}.

To obtain the $Q^2\to\infty$ expansions of the pion and $\rho$-meson form factors in
Eqs.~(\ref{int ds=Gpi0}) and (\ref{int ds=Grho0}), we use for the composite-quark wave
function (\ref{phi}) the ground-state harmonic-oscillator form
\begin{equation}
	u^{(\pi,\rho)}(k) =
	2\sqrt{\frac{1}{\sqrt{\pi}\,b_{(\pi,\rho)}^3}}
	\exp\left(-\,\frac{k^2}{2\,b_{(\pi,\rho)}^2}\right)\;.
	\label{wfHO}
\end{equation}
The parameters $b_{(\pi,\rho)}$ are chosen according to the earlier analyses
\cite{KrT01, KrP15, KrP16, KrP18}, which provided a good description of pion and
$\rho$-meson electroweak observables at low and intermediate momentum transfer. It was also
shown previously that the main results depend only weakly on the particular model wave
function used \cite{KrT09prc}.

For the integrals (\ref{int ds=Gpi0}) and (\ref{int ds=Grho0}) with the wave function
(\ref{wfHO}), the method of Ref.~\cite{KrT08} gives
	$$
	G^{(\pi,\rho)}_{ij}(Q^2)\sim
	$$
$$
	\sim\exp\left[-\,\frac{1}{2\,b_{(\pi,\rho)}^2}
	\left(\frac{M}{2}\sqrt{Q^2 + 4\,M^2} - M^2\right)\right]\times
$$
\begin{equation}
\times\sum_{k=0}^\infty\sum_{m=0}^\infty\,h^{km}_{ij}\;,
		\label{Gpi}
	\end{equation}
	$$
	h^{km}_{ij} =
	$$
	$$
	= \sum_{p=0}^{p_m}\frac{b^{2m+2k-2p}}{Q^{2k+3m-5p-1/2}}
	\frac{2^{2m+5(k+1)/2-7p}}{M^{3p-m-1/2}}C^{2m}_{4p}\frac{(4p)!}{p!m!}
	$$
	\begin{equation}
		\times\left.\frac{\partial^{2m-4p}}{\partial t^{2m-4p}}
		\left[\tilde G^{(k)}_{ij}\left(t\,,Q^2\,,\phi(t)\right)\right]\right|_{t=0}
		\;,
		\label{hkm}
	\end{equation}
$$
		\tilde G^{(k)}_{ij}(t\,,Q^2\,,t') =
$$
\begin{equation}
		= \frac{\partial^k}{\partial t'\,^k}
		\left[G^{(0J)}_{ij}(t\,,Q^2\,,t')
		\frac{\sqrt{(s-4M^2)(s'-4M^2)}}{\sqrt[4]{ss'}}\right]\;,
		\label{Gm}
	\end{equation}
where $p_m=m/2+[(-1)^m-1]/4$ and $C^{2m}_{4p}$ denotes a binomial coefficient. The
indices $i$ and $j$ take the values specified in Eqs.~(\ref{int ds=Gpi0}) and
(\ref{int ds=Grho0}); $G^{(0J)}_{ij}(t,Q^2,t')$, with $J=0$ for the pion and $J=1$ for
the $\rho$ meson, are the free two-particle form factors listed in the Appendix. The
variables $t,t'$ are related to the integration variables $s,s'$ by
	$$
	s = \frac{Q^2}{\sqrt{2}}\left(t' + t\right) + 2\,M^2 + M\sqrt{Q^2 + 4\,M^2}\;,
	$$
	\begin{equation}
		s' = \frac{Q^2}{\sqrt{2}}\left(t' - t\right) + 2\,M^2 + M\sqrt{Q^2 + 4\,M^2}\;.
		\label{ts}
	\end{equation}
The function $t'=\phi(t)$, obtained from the step function $\vartheta(s,Q^2,s')$ given in
the Appendix, specifies the boundary of the integration domain in the $(t,t')$ variables:
	$$
	t' = \phi(t) =
	$$
	\begin{equation}
		=\sqrt{1 + \frac{4M^2}{Q^2}}\left(-\frac{\sqrt{2}\,M}{Q} + \sqrt{\frac{2M^2}{Q^2} + t^2}\right)\;.
		\label{phit}
	\end{equation}
This boundary passes through $(t,t')=(0,0)$, and the neighborhood of this point gives the
dominant contribution to the asymptotic expansion.

One feature of the expansion (\ref{Gpi})--(\ref{Gm}) deserves comment. The terms
$h_{km}$ in Eq.~(\ref{Gpi}) contain not only nonnegative powers of $1/Q$ but also functions
of $1/Q$ as multiplicative factors. Consequently, each $h_{km}$ must itself be expanded in
powers of $1/Q$; in general this produces contributions with all nonnegative powers of
$1/Q$. To truncate the sums in Eq.~(\ref{Gpi}), one must therefore specify the desired
accuracy of the asymptotic expansion. Here we retain powers of $1/Q$ lower than three,
i.e., we discard terms of order $(1/Q)^3$ and higher. At this accuracy only the terms with
$k,m=0,1,2$ in Eq.~(\ref{Gpi}) contribute powers below third order. The sums in
Eq.~(\ref{Gpi}) can thus be restricted to $k,m=0,1,2$.

With this prescription, the $Q^2\to\infty$ expansions of the pion and $\rho$-meson form
factors take the following form.\\
Pion charge form factor:
$$
			G_{C}^{(\pi)}(Q^2) = G_{10}^{(\pi)}(Q^2)\sim
$$
$$
\sim\exp\left[-\,\frac{1}{2\,b_{\pi}^2}
\left(\frac{M}{2}\sqrt{Q^2 + 4\,M^2} - M^2\right)\right]\times
$$
$$
\times \left\{4\sqrt2\left( \frac{M}{Q} - \frac{3M^2}{Q^2}-\frac{9 b_\pi^2}{2Q^2}\right)\right.
\times
$$
$$
\times\left[G_E^u(Q^2)+G_E^{\bar{d}}(Q^2)\right] +
$$
	\begin{equation}
		\left.+ 32\sqrt{2}\frac{b_\pi^2}{Q^2}
		\left[G_M^u(Q^2)+G_M^{\bar{d}}(Q^2)\right]\right\}\;,
		\label{GCpi}
	\end{equation}
$\rho$-meson charge form factor:
$$
		G_{C}^{(\rho)}(Q^2) = G_{10}^{(\rho)}(Q^2)\sim
$$
$$
\sim\exp\left[-\,\frac{1}{2\,b_{\rho}^2}
\left(\frac{M}{2}\sqrt{Q^2 + 4\,M^2} - M^2\right)\right]\times
$$
$$
		\times\left\{4\sqrt2\left(\frac{M}{Q}-\dfrac{3M^2}{Q^2}-\frac{9b_\rho^2}{2Q^2}
		+\frac{64b_\rho^2M^{1/2}}{3 Q^{5/2}}\right)\right.\times
$$
$$
		\times\left[G_E^u(Q^2)+G_E^{\bar{d}}(Q^2)\right] -
$$
$$
		-4\sqrt2\left(\frac{8b_\rho^2}{Q^2}
		-\frac{64b_\rho^2M^{1/2}}{3 Q^{5/2}}\right)\times
$$
\begin{equation}
		\times\left.\left[G_M^u(Q^2)+G_M^{\bar{d}}(Q^2)\right]\right\}\;,
		\label{GCrho}
	\end{equation}
$\rho$-meson quadrupole form factor:
$$
	G_{Q}^{(\rho)}(Q^2) = \frac{2\,M_\rho^2}{Q^2}\,G_{12}^{(\rho)}(Q^2)\sim
$$
$$
\sim\frac{2\,M_\rho^2}{Q^2}\exp\left[-\,\frac{1}{2\,b_{\rho}^2}
\left(\frac{M}{2}\sqrt{Q^2 + 4\,M^2} - M^2\right)\right]\times
$$
$$
	\times\frac{32\sqrt{2}\,b_\rho^2}{Q^2}
	\left(
	\frac{2M^{1/2}}{Q^{1/2}}
	\left[G_E^u(Q^2)+G_E^{\bar{d}}(Q^2)\right]\right. +
$$
\begin{equation}
	+ \left.\left(\frac{2M^{1/2}}{Q^{1/2}} - 1\right)
	\left[G_M^u(Q^2)+G_M^{\bar{d}}(Q^2)\right]\right)\;,
		\label{GQrho}
\end{equation}
$\rho$-meson magnetic form factor:
$$
		G_{M}^{(\rho)}(Q^2) = -\,M_\rho\,G_{30}^{(\rho)}(Q^2) \sim
$$
$$
\sim\,-\,M_{\rho}\exp\left[-\,\frac{1}{2\,b_{\rho}^2}
\left(\frac{M}{2}\sqrt{Q^2 + 4\,M^2} - M^2\right)\right]\times
$$
$$
		\times\,4\sqrt2 \frac{1}{Q}\left(1 -\frac{3M}{Q}
		-\frac{9b_\rho^2}{2MQ}+\frac{24b_\rho^2}{M^{1/2}Q^{3/2}}\right)\times
$$
	\begin{equation}
		\times\left[G_M^u(Q^2)+G_M^{\bar{d}}(Q^2)\right]\:,
		\label{GMrho}
	\end{equation}

\section{Results and discussion}
	\label{sec: Sec 4}

We now discuss several consequences of the asymptotic expansions
(\ref{GCpi})--(\ref{GMrho}) derived in the preceding section. First, their origin is
manifestly relativistic. Indeed, in the nonrelativistic limit the EMFF integrals
(\ref{int ds=Gpi0}) and (\ref{int ds=Grho0}) can be evaluated explicitly, and with the
wave function (\ref{wfHO}) all form factors decrease as Gaussians with increasing momentum
transfer $Q$, as was previously demonstrated for the pion \cite{KrT09prc}. Furthermore, the
contributions of the constituent-quark magnetic form factors to the asymptotic pion and
$\rho$ charge form factors in Eqs.~(\ref{GCpi}) and (\ref{GCrho}) arise from the kinematic
relativistic effect of spin rotation (see, e.g., Refs.~\cite{KrT09, KeP91, Nov76}). The
nonzero $\rho$-meson quadrupole form factor in Eqs.~(\ref{int ds=Grho0}) and (\ref{G=f}),
and hence its asymptotic expression (\ref{GQrho}), is another manifestation of the same
effect. This can be verified by setting the spin-rotation parameters $\omega_1=\omega_2=0$
in the free two-particle quadrupole form factor entering Eq.~(\ref{int ds=Grho0}), given as
Eq.~(A3) in the Appendix. In the nonrelativistic limit, an $S$-wave $\rho$ meson has a
vanishing quadrupole form factor because its electric-charge distribution is spherically
symmetric.

Equations~(\ref{GCpi})--(\ref{GMrho}) show that the asymptotic behavior also depends on the
large-$Q^2$ behavior of the constituent-quark EMFFs. We use the form introduced in our
earlier studies \cite{KrT09, TrT13} (see also Ref.~\cite{BaY95}), which provided a good
description of light-meson electroweak properties:
$$
G^{q}_{E}(Q^2) = e_q\,f_q(Q^2)\;,
$$
\begin{equation}
	G^{q}_{M}(Q^2) = (e_q + \kappa_q)\,f_q(Q^2)\;,
	\label{qff}
\end{equation}
where $e_q$ and $\kappa_q$ are the quark charge and anomalous magnetic moment,
respectively, and
\begin{equation}
	f_q(Q^2) = \frac{1}{1 + \ln(1+ \langle r^2_q\rangle Q^2/6)}\;,
	\label{fqour}
\end{equation}
with $\langle r_q^2\rangle$ the constituent-quark mean-square radius (MSR). Thus, the quark
form factors in Eqs.~(\ref{qff}) and (\ref{fqour}) generate only weak logarithmic
corrections to the asymptotic expansions (\ref{GCpi})--(\ref{GMrho}).

We assess the accuracy of the truncated asymptotic expansions
(\ref{GCpi})--(\ref{GMrho}) by comparing them with the full form-factor expressions
(\ref{int ds=Gpi0}), (\ref{int ds=Grho0}), and (\ref{G=f}). As an illustration, we use
the pion form factor and define the ratio
\begin{equation}
	R(Q^2) = \frac{G_{Cas}^{(\pi)}(Q^2)}{G_{C}^{(\pi)}(Q^2)}\;,
	\label{R}
\end{equation}
where $G_{Cas}^{(\pi)}$ denotes the right-hand side of the truncated expansion
(\ref{GCpi}).

\begin{figure}[t]
	\centerline{\psfig{figure=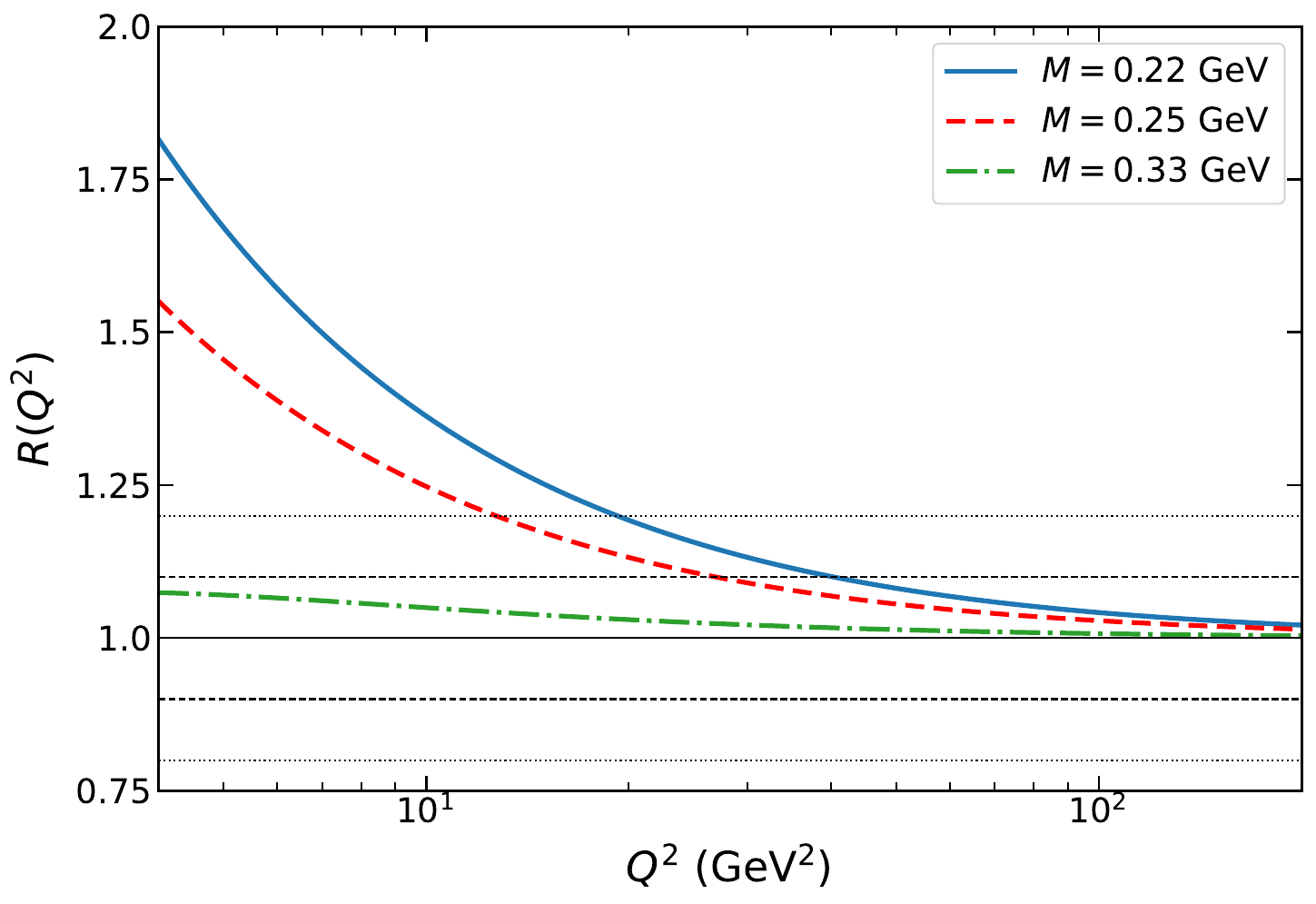,width=8.5cm}}
	\caption{Ratio of the truncated asymptotic expansion of the pion form factor,
	Eq.~(\ref{GCpi}), to the full pion form factor, Eq.~(\ref{int ds=Gpi0}), as a function
	of $Q^2$ for several constituent-quark masses. The horizontal dashed lines mark
	$\pm10\%$ deviations from $R(Q^2)=1$, while the horizontal dotted lines mark
	$\pm20\%$ deviations.}
	\label{fig:1}
\end{figure}

The ratio (\ref{R}) is shown in Fig.~\ref{fig:1}.
For the calculation of $R(Q^2)$ at finite constituent mass, the constituent-quark MSR entering
Eq.~(\ref{fqour}) was taken to be $\langle r^2_q\rangle\simeq 0.3/M^2$ \cite{KrT01}.
Figure~\ref{fig:1} shows that the truncated asymptotic expansion (\ref{GCpi}) approaches the
full pion form factor (\ref{int ds=Gpi0}) as the momentum transfer increases. The rate of
convergence depends strongly on the constituent mass. For $M=0.33$ GeV, the expansion enters
the $\pm10\%$ band around $R(Q^2)=1$ at $Q^2\simeq4$ GeV$^2$, whereas for $M=0.25$ GeV
this occurs at $Q^2\simeq13$ GeV$^2$. Including higher-order terms in Eq.~(\ref{GCpi})
would shift these values toward smaller $Q^2$.

We next compare Eqs.~(\ref{GCpi})--(\ref{GMrho}) with pQCD predictions for form factors of
zero-helicity mesons, i.e., for the invariant parts of electromagnetic-current matrix
elements between states with helicity $\lambda=0$. A zero-helicity $\rho$ meson is commonly
denoted by $\rho_L$ \cite{LeB79}. At $Q^2\to\infty$, the leading pQCD terms for the pion and
$\rho_L$ form factors are \cite{LeB79}
\begin{equation}
	F_\pi(Q^2) \sim 16\,\pi\alpha_s(Q^2)\,\frac{f_\pi^2}{Q^2}\;,
	\label{Fpi}
\end{equation}
\begin{equation}
	F_{\rho_L}(Q^2) \sim 16\,\pi\alpha_s(Q^2)\,\frac{f_\rho^2}{Q^2}\;,
	\label{FrhoL}
\end{equation}
where $\alpha_s(Q^2)$ is the running strong coupling and $f_\pi$ and $f_\rho$ are the pion
and $\rho$-meson leptonic decay constants, respectively.

The asymptotic expansions in Sec.~\ref{sec: Sec 3},
Eqs.~(\ref{int ds=Gpi0}), (\ref{int ds=Grho0}), and (\ref{G=f}), were obtained in the
canonical-spin basis rather than in the helicity basis. Therefore, to compare our IF DIR
results (\ref{GCpi})--(\ref{GMrho}) with the pQCD expressions (\ref{Fpi}) and
(\ref{FrhoL}), we need the relation between the electromagnetic form factors in the
canonical basis (the conventional Sachs form factors) and those parametrizing the current
matrix element in the helicity basis,
\begin{equation}
\langle\vec p,s,\lambda\left|j^{(\pi,\rho)}_{\mu}(0)\right|\vec p\,',s,\lambda'\rangle\;,
	\label{jlam}
\end{equation}
where $\vec p\,'$ and $\vec p$ are the particle momenta, $\lambda'$ and $\lambda$ are the
initial- and final-state meson helicities, respectively, and $s$ is the particle spin.

Canonical-spin and helicity states are related by a finite rotation (see, e.g.,
Ref.~\cite{ChS62}):
\begin{equation}
	\left.\right|\vec p,s,\lambda\rangle  =
    \sum_m\,D^{s}_{m\lambda}(\varphi,\vartheta,\psi)
	\left.\right|\vec p,s,m\rangle\;,
	\label{jlamjm}
\end{equation}
where $m$ is the spin projection on the $z$ axis and $D^{s}_{m\lambda}$ is a finite-rotation
matrix. Here $\varphi$ and $\vartheta$ are the azimuthal and polar angles of $\vec p$,
respectively; $\psi$ is an arbitrary angle whose value is immaterial for the following
calculation.

For the pion, $s=0$, so the canonical and helicity states coincide. Hence the pion form
factors in Eqs.~(\ref{int ds=Gpi0}) and (\ref{G=f}) are identical to the form factor in
Eq.~(\ref{Fpi}):
\begin{equation}
G^{(\pi)}_C(Q^2) = G^{(\pi)}_{10}(Q^2) = F_\pi(Q^2)\;.
	\label{Gpi=Fpi}
\end{equation}
The $\rho$-meson case is less direct. Using Eq.~(\ref{jlamjm}) to transform the
$\rho$-meson current matrix element (\ref{jlam}) with $s=1$, we have
$$
\langle\vec p,1,\lambda\left|j^{(\rho)}_{\mu}(0)\right|\vec p\,',1,\lambda'\rangle =
$$
$$
=  \sum_{m,m'}\,D^{1*}_{\lambda\,m}(\varphi,\vartheta,\psi)\,
D^{1}_{m'\,\lambda'}(\varphi',\vartheta',\psi')\times
$$
\begin{equation}
\times\langle\vec p,1,m\left|j^{(\rho)}_{\mu}(0)\right|\vec p\,',1,m'\rangle\;,
	\label{js1}
\end{equation}

Substituting the current parametrization (\ref{Grho}) into Eq.~(\ref{js1}) and using the
explicit rotation matrices from Ref.~\cite{Edm55}, we obtain for the longitudinal state
$\rho_L$, with $\lambda=\lambda'=0$, the following relation between the invariant form
factors in the helicity and canonical bases:
\begin{equation}
	F_{\rho_L}(Q^2) = -\left(G^{(\rho)}_C(Q^2) +
	\frac{2}{3}\,\frac{Q^2}{2M^2_\rho}G^{(\rho)}_Q(Q^2)\right)\;.
	\label{FrhoLG}
\end{equation}

Thus, the pQCD prediction (\ref{FrhoL}) applies to the form factor related to the conventional
Sachs form factors by Eq.~(\ref{FrhoLG}). Substituting Eqs.~(\ref{GCpi})--(\ref{GQrho})
into Eqs.~(\ref{Gpi=Fpi}) and (\ref{FrhoLG}) yields the $Q^2\to\infty$ expansions of
$F_\pi$ and $F_{\rho_L}$ in the helicity basis within our approach.

A useful feature of the model is the presence of parameters that control the transition from
the nonperturbative to the perturbative regime: the constituent-quark mass $M$ entering
Eqs.~(\ref{GCpi})--(\ref{GMrho}) and the constituent internal-structure parameters
$\kappa_q$ and $\langle r^2_q\rangle$ entering Eqs.~(\ref{qff}) and (\ref{fqour}). 
In the limit $M\to0$, with the scale hierarchy
\begin{equation}
\frac{M}{b}\ll \frac{b}{Q}\ll 1\;,
	\label{Mllb}
\end{equation}
our asymptotic expressions for the form factors
(\ref{Gpi=Fpi}) and (\ref{FrhoLG}) acquire the same type of structure as the corresponding
QCD results (\ref{Fpi}) and (\ref{FrhoL}). We also assume that, as $M\to0$, constituent
quarks with the internal structure described by Eqs.~(\ref{qff}) and (\ref{fqour}) become
pointlike, so that
$$
\kappa_q = 0\;,\quad \langle r^2_q\rangle = 0\;,\quad 
G_E^u(Q^2) + G_E^{\bar{d}}(Q^2) =1\;,
$$
\begin{equation}
G_M^u(Q^2) + G_M^{\bar{d}}(Q^2) = 1\;.
	\label{pointq}
\end{equation}
Under these assumptions the limit $M\to0$ gives
\begin{equation}
	F_\pi(Q^2) \sim 14\sqrt{2}\frac{b_\pi^2}{Q^2}\;,
	\label{Fpiour}
\end{equation}
\begin{equation}
	F_{\rho_L}(Q^2) \sim \frac{214}{3}\sqrt{2}\frac{b_\rho^2}{Q^2}\;,
	\label{FrhoLour}
\end{equation}
which have the same $1/Q^2$ structure as Eqs.~(\ref{Fpi}) and (\ref{FrhoL}).

In Ref.~\cite{TrT13}, using the uniform convergence of the integral
(\ref{int ds=Gpi0}) established in Ref.~\cite{KrT98}, the leading term in the simultaneous
limit $Q^2\to\infty$, $M\to0$ was obtained directly by numerical evaluation of the full
integral, without using the asymptotic expansion (\ref{Gpi}). It falls as $1/Q^2$ with a
finite coefficient and reproduces the pQCD result (\ref{Fpi}), including the coefficient of
$1/Q^2$. Thus, this independent calculation gives the same leading behavior as
Eq.~(\ref{Fpiour}).

We also note that the $M\to0$ limit of the magnetic-form-factor expansion
(\ref{GMrho}) depends on the behavior of the $\rho$-meson mass in this limit at
asymptotically large momentum transfer. This behavior can be determined by analyzing the
eigenvalue problem for the mass operator in Eq.~(\ref{complete}). Since the asymptotics of
the $\rho$-meson magnetic form factor does not affect the main conclusions of the present
work, such an analysis is beyond our scope.

Here we perform a different comparison. Lepage and Brodsky \cite{LeB79} estimated
the ratio of the asymptotic expressions (\ref{Fpi}) and (\ref{FrhoL}) to be of order 2.
Using the latest experimental values of the pion leptonic decay constant,
$f_\pi=130.2\pm1.2$ MeV, and the $\rho$-meson decay constant, $f_\rho=208-220$ MeV, with
$f_\rho$ determined from the decay $\rho^0\to e^+e^-$ under the assumption of isospin
symmetry \cite{Nav24}, we obtain
\begin{equation}
	\frac{F_{\rho_L}(Q^2)}{F_\pi(Q^2)} \sim \frac{f_\rho^2}{f_\pi^2} = 2.51 - 2.91\;,
	\label{rfrhofpi}
\end{equation}
The decay constants appearing in Eqs.~(\ref{Fpi}) and (\ref{FrhoL}) differ from the values
quoted above by a normalization factor $\sqrt{2}$; this convention difference cancels in the
ratio (\ref{rfrhofpi}). Equations~(\ref{Fpiour}) and (\ref{FrhoLour}) instead give
\begin{equation}
	\frac{F_{\rho_L}(Q^2)}{F_\pi(Q^2)} \sim \frac{b_\rho^2}{3\,b_\pi^2}\frac{107}{7}\;.
	\label{rbrhobpi}
\end{equation}

We now compare the ratios (\ref{rfrhofpi}) and (\ref{rbrhobpi}) numerically. For this purpose
we use the results of our earlier analyses \cite{KrT01, KrP15, KrP16, KrP18}, which gave a
good description of pion and $\rho$-meson electroweak data. We estimate the range of
possible values of Eq.~(\ref{rbrhobpi}) from the accuracy with which the phenomenological
parameters $b_\rho$ and $b_\pi$ are fixed.

Reference~\cite{KrP16} described simultaneously the $\rho$-meson mean-square radius (MSR)
and leptonic decay constant using a power-law wave function different from
Eq.~(\ref{wfHO}). It was shown there that $f_\rho$ depends linearly on the wave-function
parameter $b_\rho$. Thus, the experimental interval for $f_\rho$ available at that time,
with a relative uncertainty of about 5\%, corresponds to a $b_\rho$ interval of approximately
the same relative width.

We use this observation to interpret the results of Ref.~\cite{KrP15}, where the electroweak
properties of the $\rho$ meson were calculated with several model wave functions, including
Eq.~(\ref{wfHO}). That study found close results for the power-law and Gaussian wave
functions, both in good agreement with experiment. Combining Refs.~\cite{KrP15, KrP16}, we
therefore assign to the harmonic-oscillator parameter a 5\% relative interval,
$b_\rho=0.228\pm0.011$ GeV.

We apply the analogous linear dependence of the leptonic decay constant on the wave-function
parameter in our approach \cite{KrP16} to the pion analysis of Ref.~\cite{KrT01}. We use the
experimental range of $f_\pi$ available at the time of that study, while taking into account
that the value calculated in Ref.~\cite{KrT01} with the wave function (\ref{wfHO}) lies
slightly below that range. This gives $b_\pi=0.350\pm0.012$ GeV. We also note that the
results of Ref.~\cite{KrT01} with $b_\pi=0.350$ GeV provide a good description of the pion
charge-form-factor data over the full momentum-transfer range reached experimentally.

The resulting intervals of $b_\pi$ and $b_\rho$ in the harmonic-oscillator model
(\ref{wfHO}) imply
\begin{equation}
	\frac{F_{\rho_L}(Q^2)}{F_\pi(Q^2)} \sim \frac{b_\rho^2}{3\,b_\pi^2}\frac{107}{7}
	= 1.83 - 2.55\;.
	\label{rbrhobpic}
\end{equation}
Because the constituent quarks are assumed to become pointlike in the asymptotic region,
Eq.~(\ref{pointq}), the explicit form of their form factors in Eqs.~(\ref{qff}) and
(\ref{fqour}) does not affect the interval (\ref{rbrhobpic}).

The pQCD interval (\ref{rfrhofpi}) and the interval obtained in our approach
(\ref{rbrhobpic}) overlap. Thus, in the zero constituent-mass limit, the two approaches give
compatible asymptotic form-factor ratios within the adopted parameter ranges.

As noted above, Ref.~\cite{TrT13} showed by numerical evaluation of the integral
(\ref{int ds=Gpi0}) at large $Q^2$ that the pion-form-factor asymptotics in our approach
reproduces the pQCD prediction, including the coefficient multiplying $1/Q^2$. The result
obtained here for the asymptotic pion-to-$\rho_L$ relation provides a basis for an analogous
comparison involving the $\rho$ charge and quadrupole form factors through the combination
in Eq.~(\ref{FrhoLG}).

The expansions (\ref{Fpiour}) and (\ref{FrhoLour}) were obtained with the
harmonic-oscillator wave functions (\ref{wfHO}) for the relative quark motion. It was shown
in Ref.~\cite{KrT98}, however, that the leading pion-form-factor behavior $\sim1/Q^2$ in
the limit $Q^2\to\infty$, $M\to0$ holds for any wave function that ensures uniform
convergence of the integral (\ref{int ds=Gpi0}). The choice of wave function affects the
coefficient multiplying $1/Q^2$, but this coefficient remains independent of the constituent
mass. Since the integrands in Eqs.~(\ref{int ds=Gpi0}) and (\ref{int ds=Grho0}) have
closely analogous structures, the same independence of the leading power from the specific
wave-function form is expected for the $\rho$-meson form factors.

The fact that agreement with the pQCD asymptotic predictions is obtained within
instant-form dynamics may be viewed as an additional argument for its applicability to
nonperturbative problems, in line with the considerations of Ref.~\cite{LuS25}.

\section{Conclusions}
\label{sec: Sec 5}

We have studied the electromagnetic form factors of the $\pi$ and $\rho$ mesons in the
asymptotic region $Q^2\to\infty$. The analysis is based on a double integral representation
in the invariant masses of the composite system within the nonperturbative relativistic
approach developed previously by the authors, namely the modified impulse approximation 
(MIA) in instant-form dynamics. Ground-state harmonic-oscillator wave functions, 
corresponding to a quadratic confining interaction, are used. Mathematically, the problem 
reduces to the asymptotic expansion of double integrals of a special type. The expansion 
is carried out up to terms of orders less than $1/Q^3$.

The resulting power-law asymptotic behavior is intrinsically relativistic. In the
nonrelativistic limit the form-factor integrals can be evaluated analytically and exhibit a
Gaussian falloff with $Q$. In addition, the $\rho$-meson quadrupole form factor, which plays
an important role in the comparison with pQCD, is generated in the present $S$-wave model
entirely by the relativistic Wigner spin rotation and vanishes in the nonrelativistic limit.

The asymptotic expansion is obtained with the same model parameters that were fixed in the
authors' earlier description of meson electroweak properties, including electromagnetic form
factors at low and intermediate momentum transfer. No additional adjustment of the model
parameters is therefore required to obtain the asymptotic behavior. In the earlier studies,
these parameter values provided a good description of the available experimental data.

In the limit of vanishing constituent-quark mass, $M\to0$, together with the assumption that
the constituent quarks become pointlike, the asymptotic ratio of the longitudinally polarized
$\rho$-meson and pion form factors in the helicity basis is compatible with the pQCD
prediction.

Thus, the nonperturbative relativistic approach used here has the following features.

1) In the asymptotic momentum-transfer region, its results are consistent with pQCD
predictions.

2) The asymptotic ratio of the longitudinally polarized $\rho$-meson to pion
electromagnetic form factors is compatible with the pQCD expectation within the adopted 
parameter ranges without any additional fit: the asymptotic calculation uses exactly the 
same model parameters that had
already been fixed from meson electroweak properties at low and intermediate momentum
transfer. The asymptotic result is therefore a prediction of the model rather than a
consequence of tuning it to the hard regime.

3) The approach contains a parameter, the constituent-quark mass $M$, whose limiting value
describes the transition from the soft to the hard regime.

4) With the parameter values used here, the same approach has previously provided a 
successful description of available experimental data on meson electroweak properties, 
including EMFFs at low and intermediate momentum transfer.

The present results are consistent with arguments discussed in the literature in favor of
instant-form dynamics for description of the meson distribution amplitudes
at high momentum transfers in the short-distance hard-scattering mechanism \cite{LuS25}.

This work was carried out within the state assignment of Lomonosov Moscow State University.

\section*{Appendix: Free two-particle electromagnetic form factors}

For two noninteracting spin-$1/2$ fermions coupled to the pion quantum numbers
$J=S=l=0$, the free two-particle charge form factor entering
Eq.~(\ref{int ds=Gpi0}) is
$$
G^{(00)}_{10}(s,Q^2,s') =
R(s, Q^2, s')Q^2\times
$$
$$
\times\left\{(s+s'+Q^2)\left[G^u_E(Q^2) + G^{\bar d}_E(Q^2)\right]
\times\right.
$$
$$
\times\cos(\omega_1+\omega_2) +
$$
$$
+ \frac{1}{M}\xi(s,Q^2,s')\left[G^u_M(Q^2) + G^{\bar d}_M(Q^2)\right]
\times
$$
$$
\left.\times\sin(\omega_1 + \omega_2)\right\}\;,
\eqno{(A1)}
$$
For two noninteracting spin-$1/2$ fermions with the $\rho$-meson quantum numbers
$J=S=1$, $l=0$, the free two-particle electromagnetic form factors entering
Eq.~(\ref{int ds=Grho0}) are given below.

The charge form factor is
$$
G^{(01)}_{10}(s\,,Q^2\,,s') =
\frac{1}{3}R(s, Q^2, s')Q^2\times
$$
$$
\times\left\{(s+s'+Q^2)\left[G^u_E(Q^2) + G^{\bar d}_E(Q^2)\right]\times\right.
$$
$$
\times\left[2\cos(\omega_1 - \omega_2) + \cos(\omega_1+\omega_2)\right] -
$$
$$
- \frac{1}{M}\xi(s,Q^2,s')\left[G^u_M(Q^2) + G^{\bar d}_M(Q^2)\right]\times
$$
$$
\left.\times\left[2\sin(\omega_1 - \omega_2) - \sin(\omega_1 + \omega_2)\right]\right\}\;,
\eqno{(A2)}
$$
the quadrupole form factor is
$$
G^{(01)}_{12}(s\,,Q^2\,,s') =
\frac{1}{2}R(s, Q^2, s')Q^2\times
$$
$$
\times\left\{(s+s'+Q^2)\left[G^u_E(Q^2) + G^{\bar d}_E(Q^2)\right]\right.\times
$$
$$
\times\left[\cos(\omega_1 - \omega_2) - \cos(\omega_1+\omega_2)\right] -
$$
$$
- \frac{1}{M}\xi(s,Q^2,s')\left[G^u_M(Q^2) + G^{\bar d}_M(Q^2)\right]\times
$$
$$
\times\left.\left[\sin(\omega_1 - \omega_2) + \sin(\omega_1 + \omega_2)\right]\right\}\;,
\eqno{(A3)}
$$
and the magnetic form factor is
$$
G^{(01)}_{30}(s\,,Q^2\,,s') =
-2\,R(s, Q^2, s')\times
$$
$$
\times\left\{\xi(s,Q^2,s')\left[G^u_E(Q^2) + G^{\bar d}_E(Q^2)\right]\right.\times
$$
$$
\times\sin(\omega_1 - \omega_2) +
$$
$$
+ \frac{1}{4\,M}\left[G^u_M(Q^2) + G^{\bar d}_M(Q^2)\right]\left[(s+s'+Q^2)Q^2\right.\times
$$
$$
\times\left(\frac{3}{2}\cos(\omega_1 - \omega_2) + \frac{1}{2}\cos(\omega_1 + \omega_2)\right)+
$$
$$
+ \frac{1}{4}\xi(s,Q^2,s')\,
\left[\beta_1(s,Q^2,s') + \beta_1(s',Q^2,s)\right]\times
$$
$$
\times\left[\sin(\omega_1 - \omega_2) - \sin(\omega_1 + \omega_2)\right] +
$$
$$
+ \frac{1}{2}\xi^2(s,Q^2,s')\,\left[\beta_2(s,Q^2,s') + \beta_2(s',Q^2,s)\right]\times
$$
$$
\times\left.\left.\left[\cos(\omega_1 - \omega_2) + \cos(\omega_1 + \omega_2)\right]\right]\right\}\;,
\eqno{(A4)}
$$
with
$$
R(s, Q^2, s') = \frac{(s + s'+Q^2)}{2\sqrt{(s-4M^2) (s'-4M^2)}}\,
$$
$$
\times\frac{\vartheta(s,Q^2,s')}{{[\lambda(s,-Q^2,s')]}^{3/2}}
\frac {1}{\sqrt {1 + Q^2/4M^2}}\;,
$$
$$
\xi(s,Q^2,s')=\sqrt{-(M^2\lambda(s,-Q^2,s')-ss'Q^2)}\;,
$$
$$
\beta_1(s,Q^2,s') =
$$
$$
=-\,\frac{(\sqrt{s'} + 2M)(s-s'+Q^2) + (s'-s+Q^2)\sqrt{s'}}{\sqrt{s'}(\sqrt{s'} + 2M)}\;,
$$
$$
\beta_2(s,Q^2,s') = -\,\frac{1}{\sqrt{s'}(\sqrt{s'} + 2M)}\;.
$$
The angles $\omega_1$ and $\omega_2$ parameterize the Wigner rotations of the constituent
spins:
$$
\omega_1 =
\arctan\frac{\xi(s,Q^2,s')}{M\left[(\sqrt{s}+\sqrt{s'})^2 + Q^2\right] + \sqrt{ss'}(\sqrt{s} +\sqrt{s'})}\;,
$$
$$
\omega_2 = \arctan\frac{ \alpha (s,s') \xi(s,Q^2,s')} {M(s + s' + Q^2) \alpha (s,s') + \sqrt{ss'}(4M^2 + Q^2)}\;,
$$
where $\alpha(s,s')=2M+\sqrt{s}+\sqrt{s'}$. The support factor is
$\vartheta(s,Q^2,s')=\theta(s'-s_1)-\theta(s'-s_2)$, with $\theta$ the Heaviside function,
and
$$
s_{1,2}=2M^2+\frac{1}{2M^2} (2M^2+Q^2)(s-2M^2)
$$
$$
\mp \frac{1}{2M^2} \sqrt{Q^2(4M^2+Q^2)s(s-4M^2)}\;.
$$
Finally,
$$
\lambda(a,b,c) = a^2 + b^2 +c^2 - 2(ab + ac + bc)\;,
$$
$G^{u,\bar d}_{E,M}$ denote the Sachs form factors of the $u$ and $\bar d$ constituent
quarks, and $M$ is their common mass.

\end{document}